\documentclass{nature}
\usepackage[english]{babel}
\usepackage{blindtext}
\usepackage{epsfig}
\usepackage{subcaption} 
\usepackage{ae, aecompl}
\usepackage{helvet}
\usepackage{epsfig}
\usepackage{dcolumn}
\usepackage{color}
\usepackage[LGR, T1]{fontenc}
\usepackage[latin9]{inputenc}
\usepackage{amssymb}
\usepackage{amsmath}
\usepackage{mwe, tikz}
\usepackage{esint}
\usepackage{longtable}
\usepackage{url}
\usepackage{etoolbox}
\usepackage{multirow}

\newcommand\apj{Astrophys. J.,}
\newcommand\apjl{Astrophys. J. Lett.,}

\newcommand\mnras{Mon. Not. R. Astron. Soc.,}
\newcommand\nat{Nature,}
\newcommand\physrep{Phys. Rep.,}

\newcommand\aapr{Astron. Astrophys. Rev.}

\usepackage{xpatch}

\title{Astrophysics: A fast radio burst with an unexpected repeat period} 

\author{
Bing Zhang
}

\begin{document}

\maketitle

\begin{affiliations}
\item Department of Physics and Astronomy, University of Nevada, Las Vegas, NV 89154, USA. Email: zhang@physics.unlv.edu
\end{affiliations}


\begin{abstract}
Observations of millisecond-long radio bursts from beyond the Milky Way have revealed a repeat pattern with a roughly 16-day period -- a finding that adds to the enigma of the origin of these bursts. 
\end{abstract}

Mysterious flashes of radio-frequency electromagnetic radiation, which lasts for just a few milliseconds, have baffled astrophysicists since their discovery\cite{lorimer07} in 2007. Originating from outside the Milky Way Galaxy\cite{petroff19}, most of these fast radio bursts (FRBs) seem to be one-off events, but some sporadically emit repeated bursts\cite{spitler16}. On page 351, the Canadian Hydrogen Intensity Mapping Experiment fast radio bursts (CHIME/FRB) Collaboration reports\cite{periodicFRB} an intriguing regular pattern of bursts, with a period of about 16 days. 

The CHIME telescope has a large, instantaneous field of view  (about 200 square degrees) that observes light in the 400-800 megahertz frequency range, which is ideal for searching for FRBs. The FRB 180916.J0158+65 was one of the earliest repeating FRB sources discovered\cite{chime-repeaters} by CHIME. Because the source regularly falls into the telescope's field of view, it has been automatically monitored daily for an extended period of time. From 16 September 2018 to 4 February 2020, the telescope detected altogether 38 bursts from the source. Surprisingly, these bursts show a period of $16.35\pm 0.15$ days. The window of activity during each period is about 5 days, with most bursts during this window concentrated into a time of roughly 0.6 days.

Establishing a periodicity of such a long timescale for an astrophysical object is not easy, especially when only a few dozen events have been observed. One needs to carefully analyse the observational data to search for an active time window, a task that is complicated by the fact that the period of the putative regular bursts is unknown. False periods have been claimed before for other astronomical objects, such as quasars, because of the overlooked ``red noises'' -- random variations that can produce intervals of seemingly periodic behaviour\cite{vaughan16}.

The CHIME/FRB Collaboration carried out careful statistical analyses of its data, and claims that the chance of the periodicity arising from random flashes is only 1 in 10 million. There is a small possibility of `aliasing' -- the period might have been misidentified because the daily observation of the FRB source by the CHIME telescope was short. However, the authors argue that such aliasing is unlikely. Future independent confirmation of the periodicity using other telescopes would strengthen confidence in the authors' conclusion. 

Let us accept that the reported period is real. Does this help to identify the unknown mechanism that produces FRBs? Unfortunately, not really. Astronomers have been struggled to identify the astrophysical source(s) that produce FRBs, and have so far been unsuccessful. At the time of writing, 51 models of FRBs have been collected in the FRB Theory Wiki page (see go.nature.com/37acmxI)\cite{platts19}, but no `smoking-gun' observation has been made that narrows down the options. If a period was observed that is predicted by some of the models, it would provide a compelling clue, enabling us to limit the possibilities. 

For example, various models suggest that stellar remnants known as neutron stars are emitters of FRBs -- either magnetically powered neutron stars\cite{popov07} or those powered by loss of rotational energy of the star\cite{cordes16}. The analogues of these objects in our Galaxy are called magnetars and radio pulsars, respectively, and spin with a period of the order of seconds or sub-seconds\cite{kaspi10}. The identification of a seconds-long period from an FRB source would therefore immediately reveal it to be a neutron star. This happened for a less-spectacular type of sporadic radio burst in our Galaxy, the rotating radio transients\cite{mclaughlin06}. 

However, searches for short periodicities from repeating FRBs have so far been fruitless\cite{zhangy18}. The approximately 16-day period of FRB 180916.J0158+65 is too long to be the period of a spinning neutron star. Indeed, such a long period was not predicted by any FRB theory before this discovery. 

The discovery therefore stirred up intense brain-storming within the CHIME/FRB Collaboration, and by other scientists studying FRBs, when the results were first released. One possibility considered by different groups is that the FRB source is in a binary system involving a neutron star, and that the approximately 16-day period is the orbital period of that system\cite{periodicFRB,ioka20,lyutikov20}. More specifically, it has been speculated\cite{zhang17} that FRBs could be produced by direct interactions between an astronomical stream of particles, such as an intense stellar wind produced by a massive star, and the magnetosphere around a neutron star. 

But if the companion of the neutron star is a massive star, the two stars would need to be separated by about a quarter of the distance between Earth and the Sun to produce the period reported by the CHIME/FRB Collaboration\cite{ioka20,lyutikov20}. This is too far apart for such a direct-interaction scenario to work. It therefore seems that a binary system might not explain how the radio bursts of FRB 180916.J0158+65 are produced (although one model\cite{ioka20} suggests that an aurora-like inflow of the particles from a companion to a neutron star is essential for driving the emissions). However, the periodicity of the emissions can be interpreted as a consequence of there being an FRB source in the binary: the stellar wind of the neutron star might open up a `window' in the otherwise radio-obscuring stellar wind of the companion star, allowing the FRBs to escape\cite{ioka20,lyutikov20} (Fig.1a). This window is observed periodically from Earth as the binary system rotates. 

A second possibility discussed by several groups\cite{periodicFRB,levin20,zanazzi20,yang20} is that the FRB-generating neutron star is deformed, and that its emission region precesses like a gyroscope (Fig.1b). In this picture, the neutron star's spin period is much shorter than 16 days, but its FRB emission is focused into a narrow beam -- and this beam sweeps Earth about every 16 days, generating the observed period. The precession could be spontaneous\cite{levin20,zanazzi20}, or it could be induced by a companion in a close binary system\cite{yang20}. However, such precession is probably not necessary to produce FRBs in the first place. 

Finally, one can argue that the roughly 16-day period is that of an extremely slow magnetar\cite{beniamini20}. This is, however, quite a stretch -- it is unclear whether such slow magnetars exist, and, if they do, whether they can generate actively repeating radio bursts. Overall, scientists will need to expend some efforts to accommodate the reported period of FRB 180916.J0158+65 in their models.

Future continued monitoring of this and other repeating FRB sources is essential for solving the mystery of the unexpected period. One can imagine three possible outcomes. First, after long-term monitoring, bursts show up outside the active window reported by the CHIME/FRB Collaboration for this source. If so, the supposed periodicity would disappear -- and at least some theorists would breathe a sigh of relief. The second possibility is that long-term monitoring validates the claimed period of the bursts for this source, but that no other FRB source displays a clear long-term periodicity. This periodicity of FRB 180916.J0158+65 can be then understood as a peculiarity of that system, and not as something that is intrinsic to FRB production in general.

But the final possibility is the most intriguing: that long-term periodicity is the norm for repeating FRBs. If so, then such periodicity might be at the heart of the FRB mechanism -- and it would mean that these natural phenomena are defeating the ability of the human imagination to explain it. More creative ideas would be needed to identify the missing link between theory and observation.



\newpage

\begin{figure}[t]
\begin{tabular}{c}
\includegraphics[keepaspectratio, clip, width=1\textwidth]{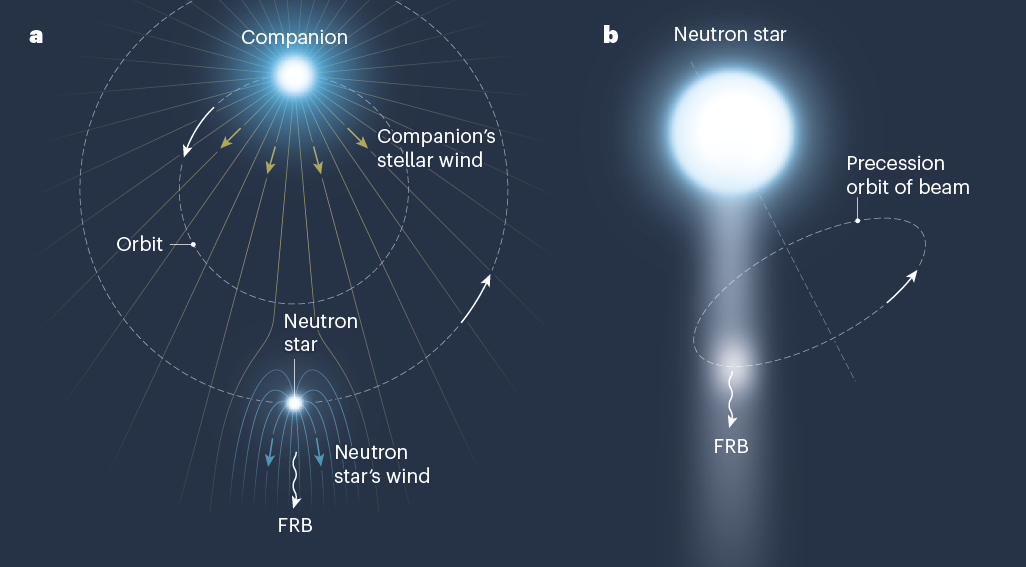} \\
\end{tabular}
\caption{{\bf Two possible scenarios to explain the observed periodicity of a fast radio burst (FRB)}. The CHIME/FRB Collaboration report\cite{periodicFRB} that FRBs from a source called FRB 180916.J0158+65 repeat with a period of about 16 days. {\bf a,} The source might be a neutron star in a binary system with a massive companion star. The companion produces a strong `wind' of particles that could obscure radio waves from the neutron star. But if the neutron star has its own stellar wind, this could deflect the companion star's particle flow, opening up a window behind the neutron star from which FRBs can escape. These FRBs could be observed when the window orbits through Earth's field of view. {\bf b,} Another scenario is that the FRBs are emitted in focused `beams' from the magnetosphere of a highly magnetized neutron star, or from regions far beyond the magnetosphere (the exact region of FRB emission is not shown here, for simplicity). These beams precess like a gyroscope, periodically entering Earth's field of view.
}
\label{fig:lc_sp}
\end{figure}

\end{document}